\documentclass[10pt,lettert,twoclumn]{IEEEtran}
\usepackage{graphicx}
\usepackage{cite}
\usepackage{color}
\usepackage{amsfonts}
\usepackage{amsmath}
\usepackage{subfigure}

\title{Buffer Aided Relaying Improves Both Throughput and End-to-End Delay} 
\author{Javad~Hajipour$^1$, Amr Mohamed$^2$, and Victor~C.~M.~Leung$^1$\\ 
$^1$\{hajipour,~vleung\}@ece.ubc.ca, $^2$\{amrm\}@ieee.org}
\begin{document}
\maketitle
\IEEEpeerreviewmaketitle
\graphicspath{{./fig/}}
\begin{abstract}
Buffer aided relaying has recently attracted a lot of attention due to the improvement in the system throughput. However, a side effect usually deemed is that buffering at relay nodes results in the increase in packet delays. In this paper, we study the effect of buffering relays on the end-to-end delay of users' data, from the time they arrive at source until delivery to the destination. We use simple discussions to provide an insight on the overall waiting time of the packets in the system. By studying the Bernoulli distributed channel conditions, and using intuitive generalizations, we conclude that the use of buffers at relays improves not only throughput, but ironically the end-to-end delay as well. Computer simulations in the settings of practical systems confirm the above results.
\end{abstract}

 \vspace*{-4mm}
\section{Introduction}\label{sec:intro}

Wireless relays are promising solutions for enhancing the capacity and coverage of cellular networks. Usually in the literature in this area, it is assumed that the relaying is performed in two consecutive subslots of a transmission interval; i.e., in the first subslot, the base station (BS) transmits to the relay and in the second one, the relay forwards the received data to the mobile terminal. Recently it has been shown that using the buffering technique at relays can improve the system throughput~\cite{JR:buffering3N,JR:perf_rateless,JR:buf_fixedmixed,JR:buf_chal}. 
This is achieved due to the fact that the buffering capability allows the relay to store the packets when the channel condition is bad and transmit when it is good. The drawback for this capability is usually deemed to be the increase in the packet delays due to queueing in the relay, and the works in~\cite{JR:buffering3N,JR:perf_rateless,JR:buf_fixedmixed,JR:buf_chal} have tried to investigate and discuss the trade off between throughput and delay. This is however based on the assumption of infinitely backlogged buffers in the source (i.e., BS), and considers the queueing delay only in the relay buffer without taking into account the queue dynamics at the BS.

In this letter, we take into account the queue dynamics both in the source and the relay and study the effect of buffering in relay, on the end-to-end delay that data bits experience since their arrival at the source until delivering to the user. For this, we provide simple reasoning and discuss the cause of queue formation in a simple queueing system, and based on that investigate the delay performance in buffer aided relaying. We conclude that the buffering relays in fact improve throughput as well as the end-to-end delay, and using simulations, we verify the discussed perspective. To the best of our knowledge, this is the first work that discusses the effect of buffering relays on the overall waiting time in a relaying network and provides the above conclusion and insight.

\vspace*{-4mm}
\section{Effect of Buffer Aided Relaying On The End-to-End Delay}\label{sec:delimp}
In this section, first we study a simple queueing system, and then based on that investigate the delay performance in a three node relay network with simple data arrival and service processes. Then we discuss general cases and present the intuitions about the end-to-end delay performance.

\vspace*{-4mm}
\subsection{Simple Queueing System}\label{sec:simpq}
\begin{figure}[!t]
\centering
\includegraphics[width=0.3\paperwidth]{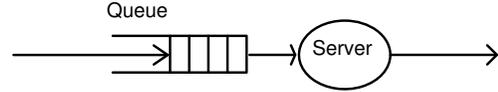}%[width=.7 \columnwidth,height=0.5 \columnwidth]{queue.eps}
\caption{Simple queueing system}\label{fig:queue}
\end{figure}

Consider a single buffer which is fed by a deterministic data arrival process and served by a single server, as shown in Figure~\ref{fig:queue}. We assume that time is divided into slots with equal lengths, which are indexed as $t=1, 2, ...$. The source data size is $N$ bits and starting from $t=1$, one bit arrives in the buffer. In particular, the first bit arrives at $t=1$ and the last one arrives at $t=N$. For simplicity, we assume that the arrivals occur at the beginning of time slots. On the other hand, the server can serve only one bit per slot, where the service is in fact to deliver the data bits to the destination. If the server is not busy when a bit arrives in the buffer, it will serve the arrived bit; otherwise the arrived bit will be queued until the server finishes its current service. Therefore, the overall waiting time for each bit is composed of the waiting time in the queue and the service time in the server.

We note that if the server is active in each of the time slots $t=1, ..., N$, each bit will be served immediately after its arrival. In this case, there is no queue formed in the buffer and consequently, each bit experiences an overall waiting time equal to one time slot, which is due to the time spent in the server. Accordingly, the data bits will arrive in the destination at the beginning of time slots $t=2, ..., N+1$. However, if the server is inactive in the first time slot, the first bit has to wait in the buffer until time slot $2$, to be served. Then in time slot $2$, when the second bit arrives, the server is busy with serving the first bit. Therefore the second bit also experiences one slot waiting in the queue and one slot in the server and similar delays happen for all the following bits. In other words, the delay in the starting time of the server causes the nonzero queueing delay for the first bit, which is transferred to the subsequent bits as well.

\begin{figure*}[!t]
 \centerline{
 \subfigure[]
 {\includegraphics[width=0.23\paperwidth]{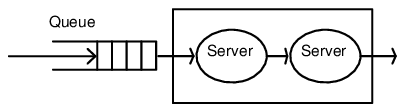}
 \label{fig:prompt}
 }
 \vspace*{-2mm}
\hspace*{1mm}
 \subfigure[]
 {\includegraphics[width=0.3\paperwidth]{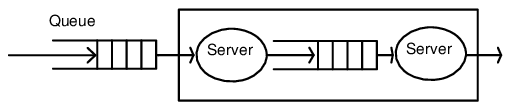}
 \label{fig:nonprompt}
}
\vspace*{-2mm}
\hspace*{1mm}
\subfigure[]
{
\begin{tabular}{l l}            % centered columns (4 columns)
\hline\hline                        %inserts double horizontal lines
$s_1s_2$ & probability \\ [0.5ex]   % inserts table
\hline                              % inserts single horizontal line
GG & $p_1p_2$\\
GB & $p_1(1-p_2)$\\
BG & $(1-p_1)p_2$\\
BB & $(1-p_1)(1-p_2)$\\
%heading
\hline                              %inserts single line
\end{tabular}
\label{fig:chstates}
}
}
\caption{(a) Conventional relaying system  (b) buffer aided relaying system (c) joint channel states }
 \label{fig:tandem}
 %\vspace*{-4mm}
 \end{figure*}

Based on the above, we note that if the server gets inactive in time slot $x \in \{1, ..., N\}$, it adds one slot to the queueing delay (and the overall waiting time) of every bit arrived in slot $x$ or afterward. In general, the packet arrived in time slot $i$ will experience a queueing delay of $n_i$ and will be delivered at slot $i+n_i+1$, where $n_i$ indicates the number of slots before and including $i$, in which the server was inactive.

Considering the above discussions, it is clear that the cause of queue formation in the studied system is the interruption in the operation of the server, which is translated to the queueing delays for the data bits.

\vspace*{-4mm}
\subsection{Relaying System with Simple Data Arrival and Service Time}\label{sec:relaysys_sim}
Now consider a relaying network, with one source node, i.e., the BS, one relay node and one destination (or user) node, where the relay works based on Decode and Forward (DF) technique. It is assumed that there is no direct link between BS and the user, and the transmissions are done only through the relay. There is only one channel in the system, which can be used for either transmissions from BS to relay or from relay to user. Each time slot is divided into two subslots, where BS and relay can transmit in the first and second subslot, respectively. We use $s_1$ and $s_2$ to indicate the BS channel condition (for the link between BS and relay) and relay channel condition (for the link between relay and user), respectively. These variables can either be ``Good'' or ``Bad'' , meaning respectively that it is possible to transmit one or zero bit on the corresponding channel.  The probability of being ``Good`` is respectively $p_1$ and $p_2$ for BS and relay channel. Figure~\ref{fig:prompt} shows a system model with conventional relaying, where the relay does not have buffer and therefore it has to transmit its received data immediately in the next subslot. The server 1 and server 2 indicate respectively the wireless channel from BS to relay and from relay to user. On the other hand, Figure~\ref{fig:nonprompt} indicates a relaying network, where the relay is equipped with a buffer which allows it to store the data bits and transmit whenever its channel is good. In both of the figures, the rectangle enclosed around the servers is to abstract the overall serving behavior of the system from the time that BS starts to transmit data bits until their delivery to the user. 
Note that the works in~\cite{JR:buffering3N,JR:perf_rateless,JR:buf_fixedmixed,JR:buf_chal}, in fact study the delay by considering only the time a packet spends inside this rectangle and do not take into account the waiting time in the BS queue, which occurs before the transmission from the BS to the relay.  

In the following, we consider the data arrivals in the BS buffer as the deterministic process, with $N$ bits, mentioned in the previous subsection. Taking into account the overall service behavior of the systems, we discuss the overall waiting time of data bits in both conventional and buffer aided relaying systems. The overall waiting time is in fact the end-to-end delay, from the time that a data bit arrives in the BS buffer until it is delivered to the user. 

Figure~\ref{fig:chstates} shows the different states for the joint conditions of BS and relay channels, in which $G$ and $B$ indicate ``Good'' and ``Bad'' respectively. We note that the system with conventional relaying serves the data bits only when $s_1s_2=GG$, and with the probability of $p_t=p_1p_2$. In the other three cases, i.e. when either or both of $s_1$ and $s_2$ are ``Bad'', the data bits remain in the BS buffer and are not transmitted. Therefore, based on the discussions in the previous subsection, the overall server in the system is inactive with probability 
\begin{equation}
q^{nb}_t=P(GB)+P(BG)+P(BB)=1-p_t=1-p_1p_2
\end{equation}
where $q^{nb}_t$ indicates the interruption probability for the overall server, in the system without buffering in the relay. Considering this, in each time slot, the probability of ``increase of one slot'' in the overall waiting time of data bits present in that time slot or arrived after that, is $q^{nb}_t=1-p_1p_2$. Here, the increase in the overall waiting time is due to the increase in the BS queueing delay of those bits.

Now consider the system with buffering relay. We note that if the channel conditions are as $BB$ in time slot $x$, similar to the system with conventional relaying, there will be an increase of one slot in the overall waiting time of the packets present in the time slot $x$ or arriving afterward. However for the channel conditions as $GB$ and $BG$, the case is different. In order to clearly investigate these states, first we consider the following example:
\begin{itemize}
  \item In time slot $t=1$, the channel conditions are as $GB$. Therefore, in the first subslot, the data bit $1$ will be transmitted from BS to relay; but due to the ``Bad'' channel condition of relay, it will not be transmitted to the user in the second subslot and will be stored in the relay's buffer.
\item In time slot $t=2$, the channel conditions are as $BG$. Therefore, in the first subslot, there will not be any transmission from BS to relay and the overall waiting time of data bits $2, ..., N$ will be increased \emph{one} slot. However, due to good condition of relay channel, the data bit $1$ will be transmitted from relay's buffer to the user, in the second subslot.
\end{itemize}
In the above example, it is observed that the data bit $1$ is served by the relay in time slot $t=2$ and therefore it is delivered to the user at time slot $t=3$. This has become possible due to the queueing of that bit in the relay's buffer. Note that with conventional relaying, however, in the above example, the data bit $1$ would remain in the BS queue in both time slot $t=1$ and $t=2$ and the overall waiting time would increase \emph{two} slots for all the data bits.
Based on the above discussion and considering the nonzero probability of having channel conditions as $GB$ and $BG$ in two consecutive time slots, it can be concluded that $q^{b}_t<q^{nb}_t$, where $q^{b}_t$ is the interruption probability for the overall server in the system with buffering relay. In other words, the buffering capability in relay improves the overall waiting time for data bits. This is achieved due to the fact that the queue size in the BS is reduced and the data bits transferred to the relay buffer, enable the use of relay channel efficiently.

\vspace*{-4mm}
\subsection{General Relaying System}\label{sec:relaysys}
Now consider a general scenario, with general distributions for data arrival and channel condition processes.
We use $r_{br}(t)$, $r_{ru}(t)$, $r_{bu}(t)$ to show respectively, the achievable transmission rate at time slot $t$ between BS and relay, relay and user and BS and user. Without buffering, the BS needs to transmit to the relay in the first subslot and then the relay has to forward it immediately in the next subslot. We know that in this case, the end to end achievable rate between BS and the user is $r_{bu}(t)=\frac{1}{2}\min\{r_{br}(t), r_{ru}(t)\}$. Due to this, the transmission in each slot, is limited by the link with the worst channel condition in that time slot. 

However when buffering is used at the relay, this limitation is relaxed and therefore BS has the opportunity to transmit continuously to the relay when the channel condition from BS to relay is good. Then the relay can buffer them to transmit when the channel from relay to user is good, and there is no necessity for the immediate forwarding of the data.

We note that when buffering is employed in the relay, there needs to be a scheduling policy, to decide in each subslot on allocating the channel to BS or relay transmissions, such that the queues in BS and relay remain stable i.e. their queue sizes stay bounded. For this, the well-known Max-Weight (MW) algorithm can be used, which has the largest stability region, i.e., equal to the system capacity region~\cite{book:stochasticopt}. 

Recall that the buffering improves the system throughput~\cite{JR:buffering3N,JR:perf_rateless,JR:buf_fixedmixed,JR:buf_chal}. Improvement in the throughput, is equivalent to the improvement in the end-to-end service rate of the data arrived in BS buffer. In other words, the increase in the system throughput, means that more data is transferred from BS to the user, or equivalently, the same data is transferred from BS to the user, in a less time, and therefore the end-to-end delay is reduced.

Based on the above discussion, we make the following conclusion: Although buffer aided relaying results in queueing delay on the data arrived in relay, it also facilitates data transfer from BS to the user and leads to a large reduction in queueing delay in BS; therefore the overall effect is the improvement in the end-to-end delay. In summary, we state this as follows:

\textbf{Proposition:} Using buffer at relay, improves the system throughput and therefore it reduces the end-to-end delay.

\textbf{Remark:} Note that the cost of these improvements is the memory needed for buffering as well as the need for a scheduling algorithm to keep the queues stable.

\vspace*{-4mm}
\section{Numerical Results}\label{sec:eval}
To verify the presented discussion, we have conducted extensive Matlab simulations over 10000 time slots. Simulation parameters are selected according to table \ref{table:simpar}. It is assumed that the channel fading is flat over the system bandwidth and constant during each time slot, but it can vary from one slot to another. For the link between relay and user, Rayleigh channel model is used and for the link from BS to relay, Rician channel model with $\kappa$ factor equal to 6 dB. In the case of conventional relaying, BS transmission and relay transmission are done in consecutive subslots. For buffer aided relaying, we have used MW policy to decide about the scheduling of transmission in each subslot, either from BS or relay buffer.

\renewcommand{\arraystretch}{.6}
\begin{table}[!t]
\caption{Simulation Parameters}   % title of Table
\centering                          % used for centering table
\begin{tabular}{l l}            % centered columns (4 columns)
\hline\hline                        %inserts double horizontal lines
Parameter Name & Setting \\ [0.5ex]   % inserts table
%heading
\hline                              % inserts single horizontal line
Cell Radius & 1000m \\
Min UE-BS distance & 50m \\
BS Antenna Height & 15m\\
Relay Antenna Height & 10m\\
User Antenna Height & 1.5m\\
Relay Distance from BS & 1/2 cell radius\\
Pathloss Model & From \cite{simulation}\\
Channel Bandwidth & 180 KHz\\
Time Slot Duration & 1ms\\
Noise Power Spectral Density  & -174dBm/Hz \\
Traffic Model & Poisson \\
Packet Size & 1Kb\\
\hline                              %inserts single line
\end{tabular}
\label{table:simpar}          % is used to refer this table in the text
\end{table}

\begin{figure*}[!t]
 \centerline{
 \subfigure[]
 {\includegraphics[width=0.3\paperwidth]{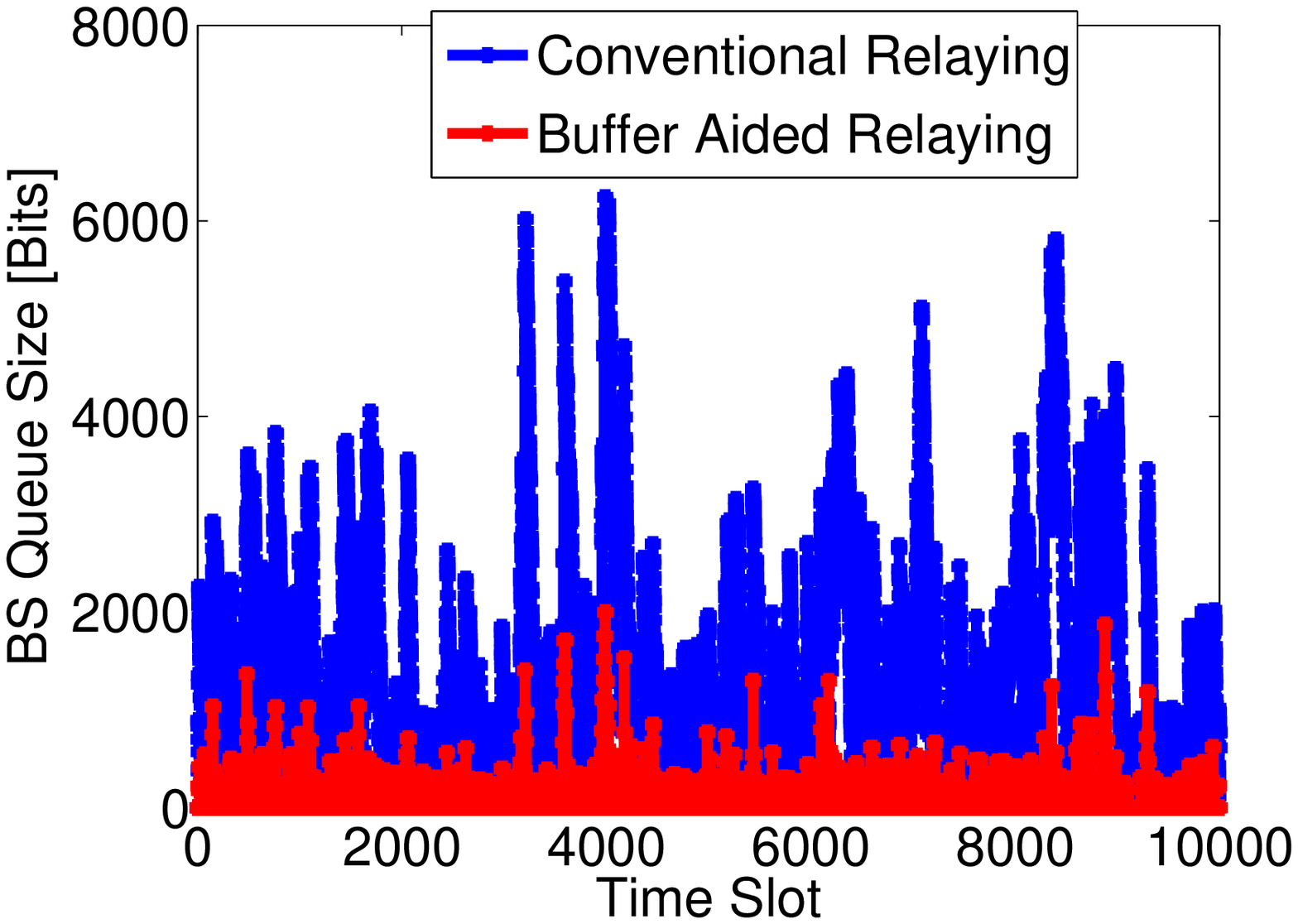}
 \label{fig:BS_50}
 }
 \vspace*{-2mm}
 \subfigure[]
 {\includegraphics[width=0.3\paperwidth]{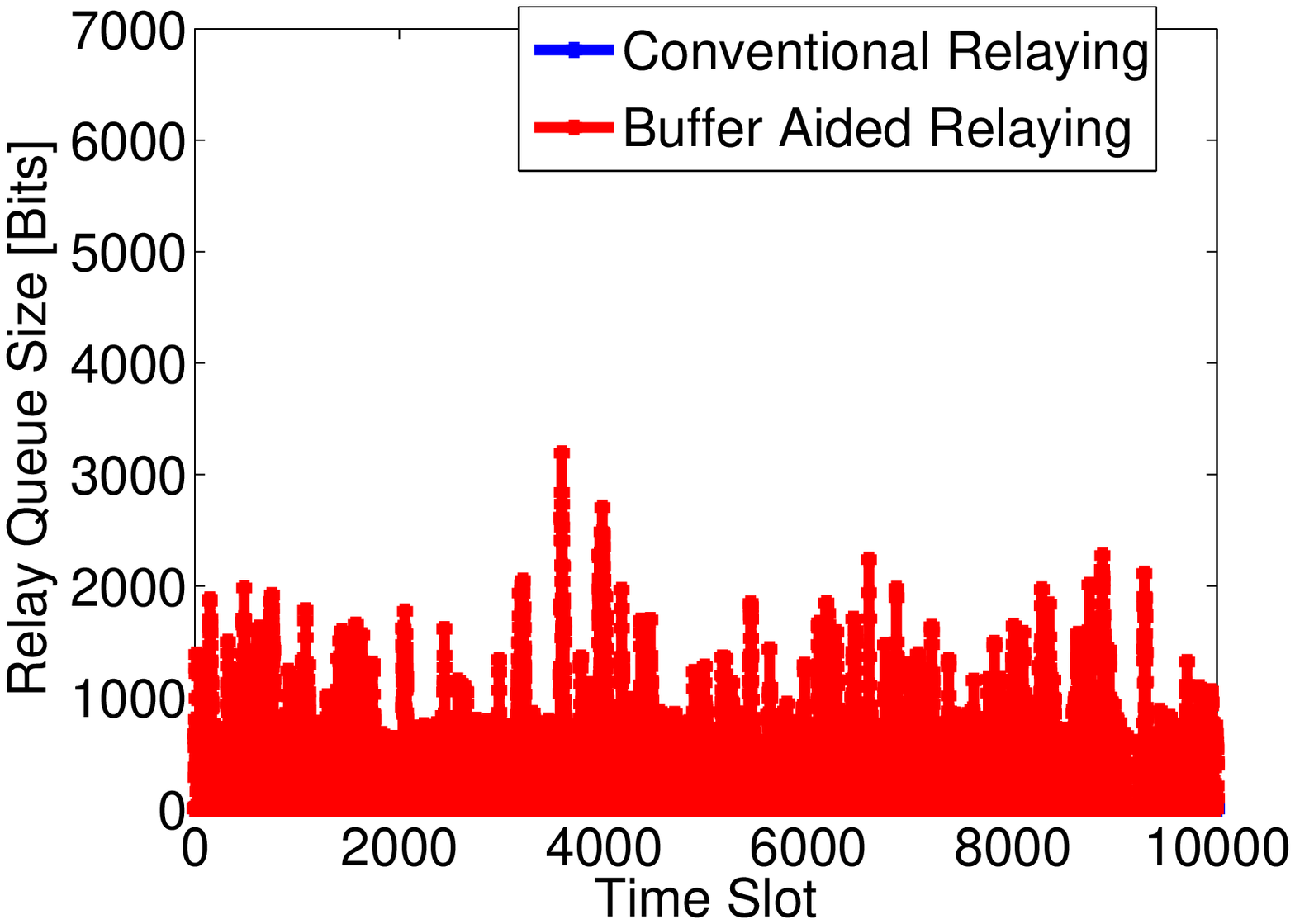}
 \label{fig:RS_50}
}
\vspace*{-2mm}
 \subfigure[]
 {\includegraphics[width=0.3\paperwidth]{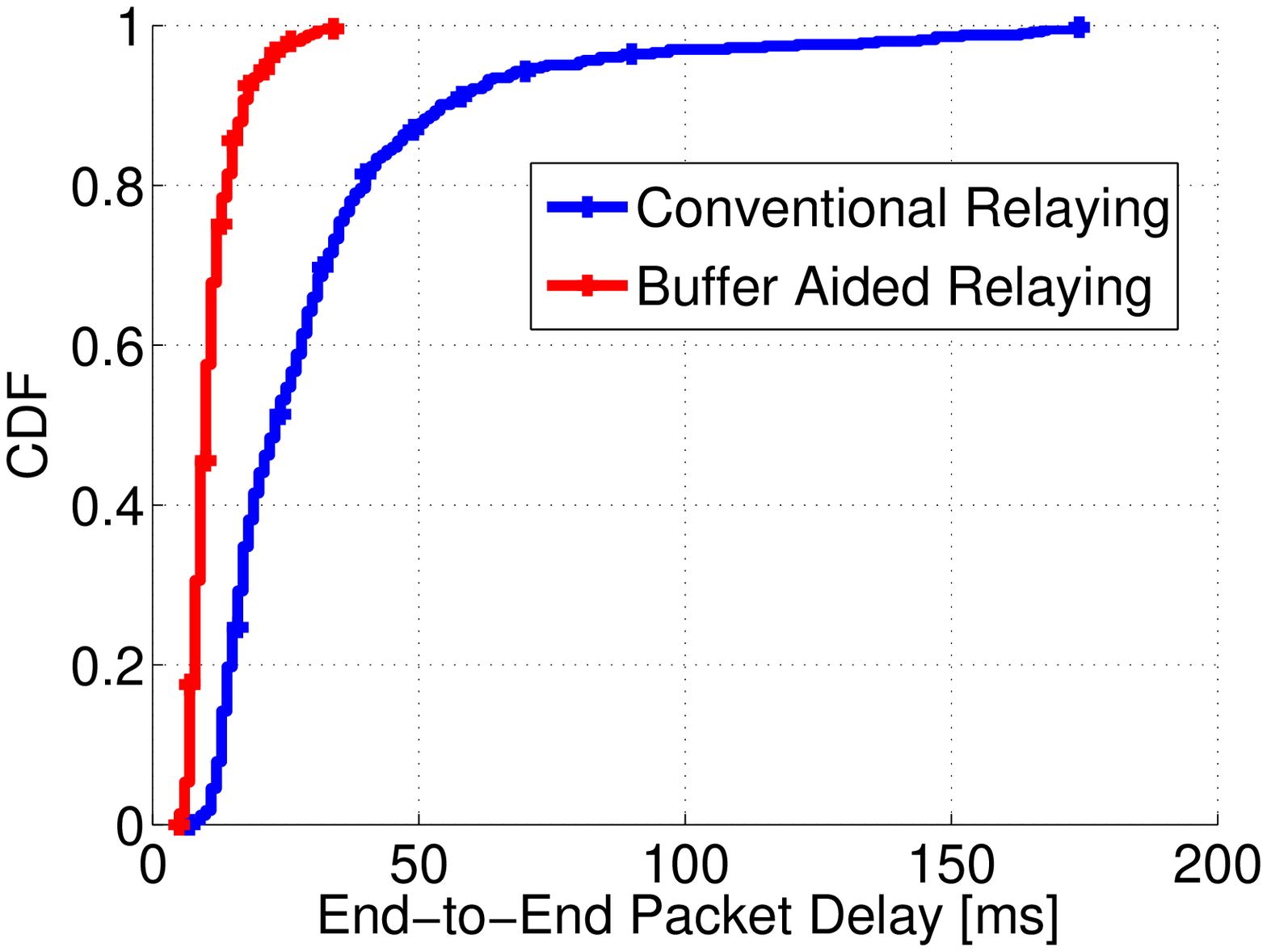}
 \label{fig:cdf_del_50}
}
}
\caption{(a) BS queue size over time (b) relay queue size over time (c) CDF of end-to-end packet delays; at the arrival rate of 50 packets/slot.}
 \label{fig:arr50}
 \vspace*{-2mm}
 \end{figure*}

\begin{figure*}[!t]
 \centerline{
\subfigure[]
  {\includegraphics[width=0.3\paperwidth]{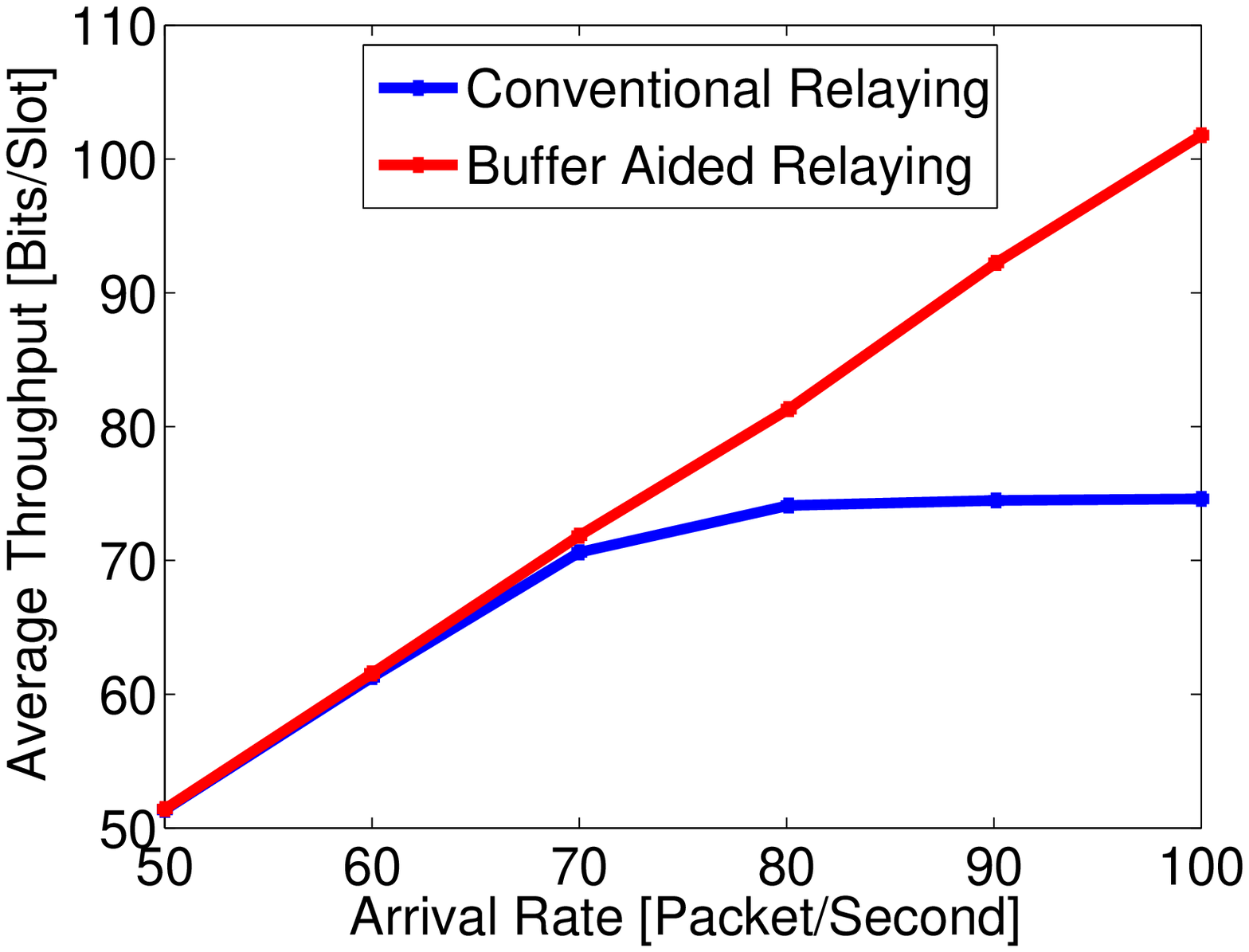}
 \label{fig:vsar_thpt}
}
% \vspace*{-2mm}
 \subfigure[]
{\includegraphics[width=0.3\paperwidth]{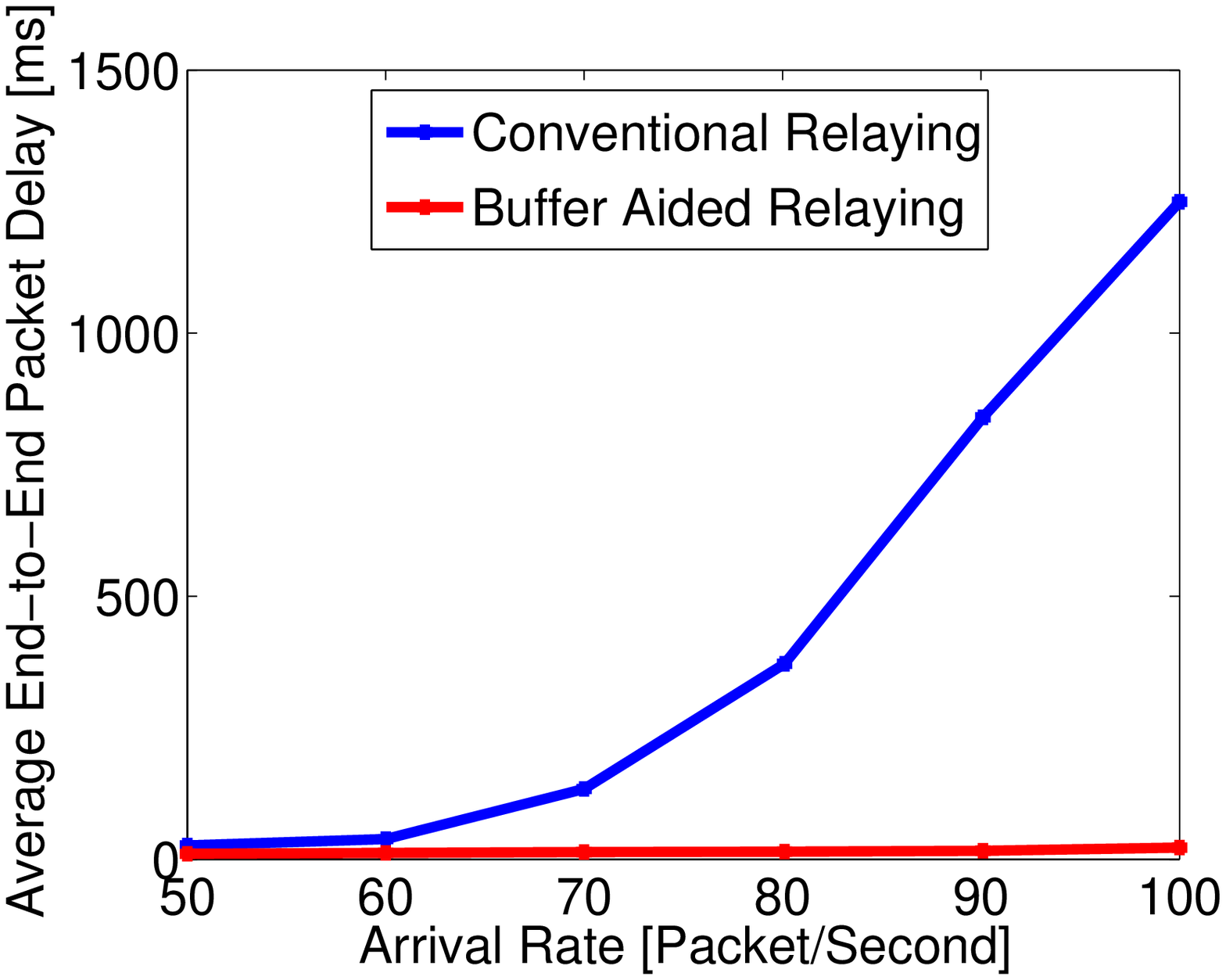}
 \label{fig:vsar_del}
}
}
\caption{Effect of packet arrival rate on (a) average throughput in each time slot (b) average end-to-end packet delay.}
 \label{fig:vsar}
 %\vspace*{-4mm}
 \end{figure*}

\begin{figure*}[!t]
 \centerline{
 \subfigure[]
 {\includegraphics[width=0.3\paperwidth]{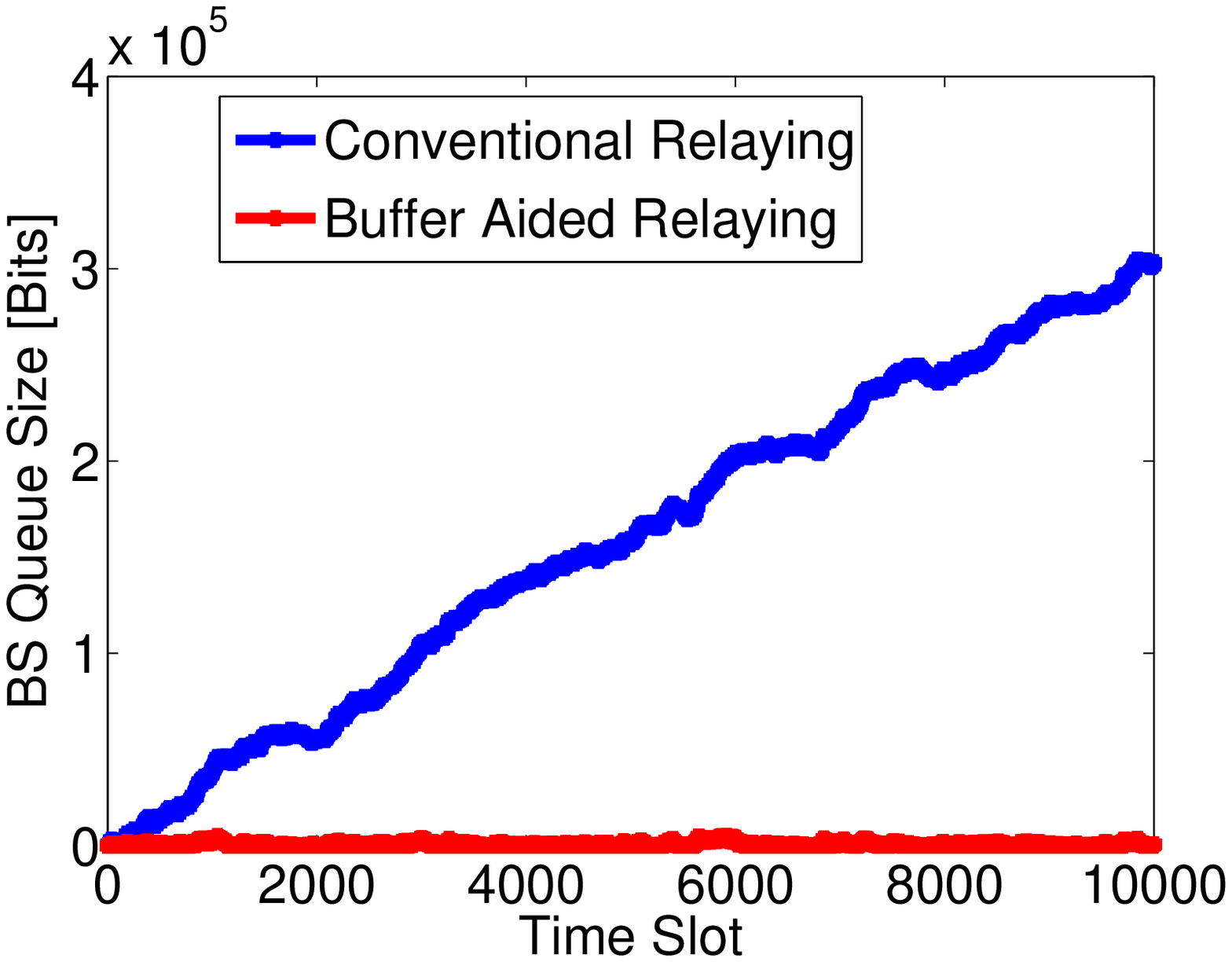}
 \label{fig:BS_100}
 }
%  \vspace*{-2mm}
 \subfigure[]
 {\includegraphics[width=0.3\paperwidth]{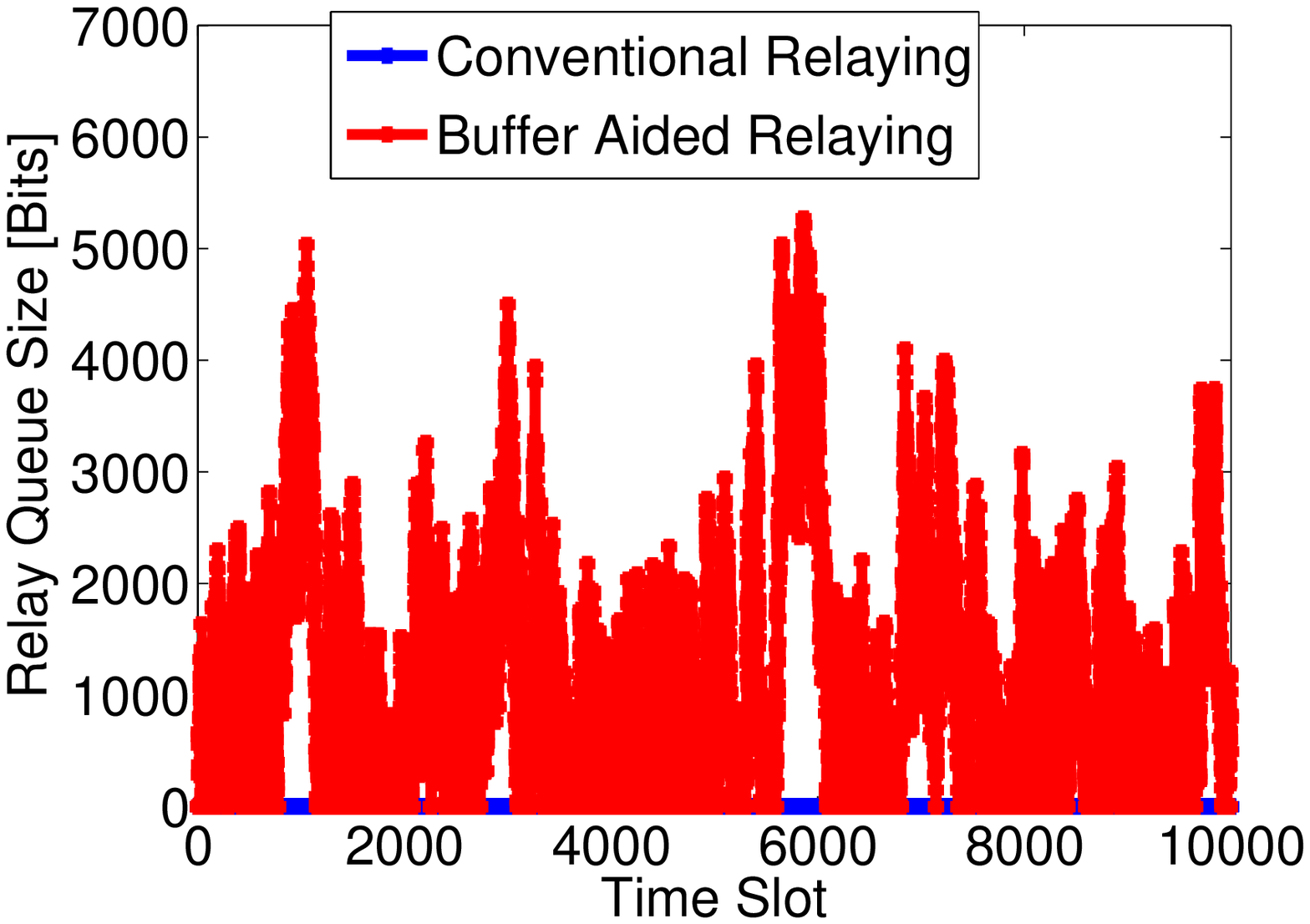}
 \label{fig:RS_100}
}
% \vspace*{-2mm}
 \subfigure[]
 {\includegraphics[width=0.3\paperwidth]{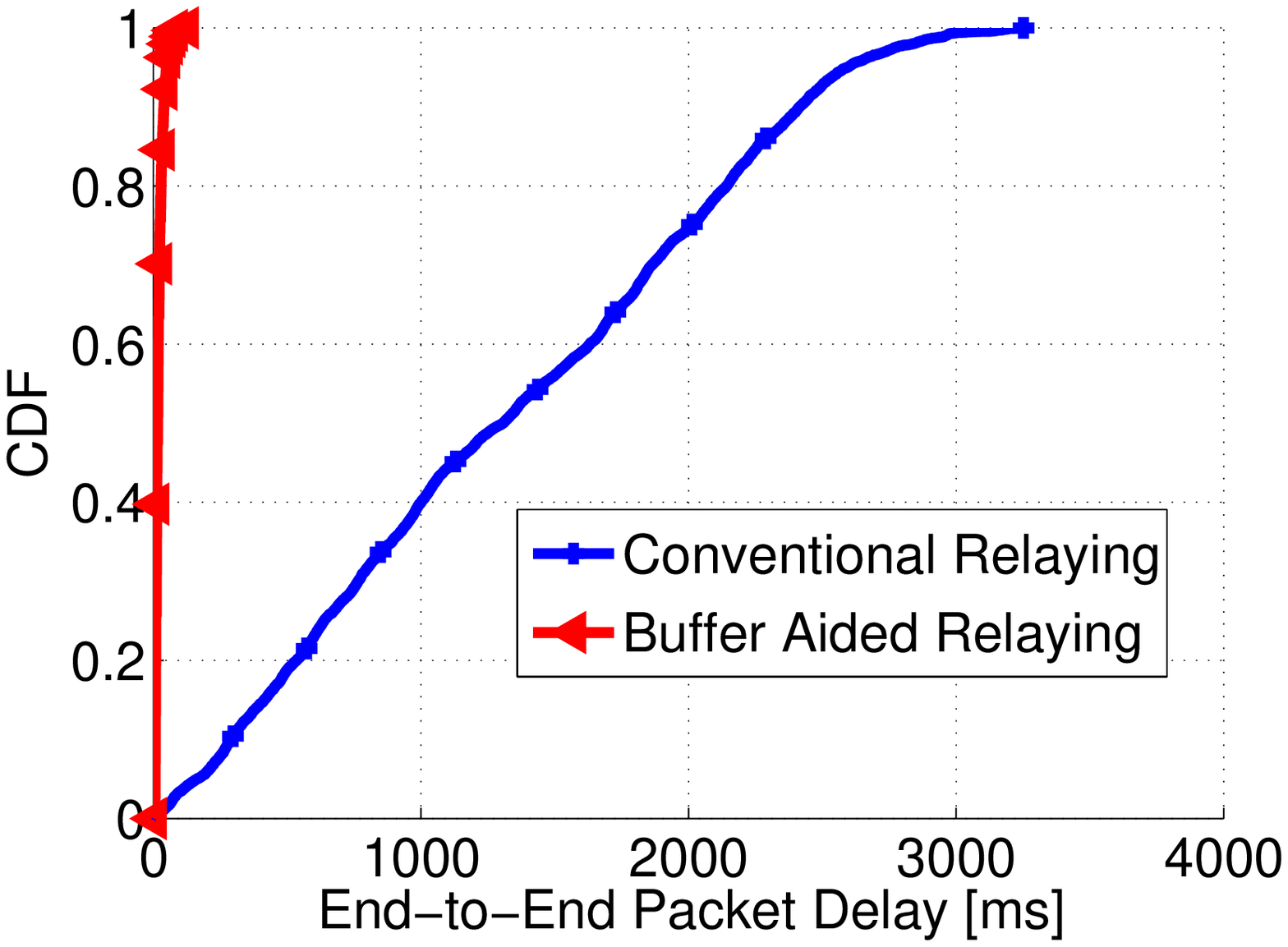}
 \label{fig:cdf_del_100}
}
}
\caption{(a) BS queue size over time (b) relay queue size over time (c) CDF of end-to-end packet delays; at the arrival rate of 100 packets/slot.}
 \label{fig:arr100}
 %\vspace*{-4mm}
 \end{figure*}

Figure~\ref{fig:arr50} shows the BS and relay queue sizes over time, and the cumulative distribution function (CDF) for end-to-end packet delays, at the arrival rate of 50 packets/second. It is observed that with buffer aided relaying, although data are queued in RS, the BS queue size in each time slot is reduced significantly. This results in lower end-to-end packet delays in buffer aided relaying compared to conventional relaying, as shown in Figure~\ref{fig:cdf_del_50}. In particular, in this scenario, the average end-to-end packet delays are 12 ms and 38 ms, respectively in buffer aided and conventional relaying.

Figure~\ref{fig:vsar} displays the effect of increase in packet arrival rate on the throughput and delay performances. It is observed that up to the arrival rate of 60 packets/second, conventional relaying is able to serve the arrived data and results in the same amount of throughput for the user. However after that, due to low capacity, it starts to get saturated which leads to queue instability and large end-to-end delays for packets. In contrast, buffer aided relaying is able to provide the throughput equal to the data arrival rate, in all packet arrival rates, and therefore leads to very low end-to-end packet delays. 

In order to have a complete picture, we also present the system performance at the arrival rate of 100 packets/second. Figure~\ref{fig:arr100} shows that in conventional relaying, the BS queue grows unbounded; this is due to the low capacity of relaying channel which is unable to serve all the arrived data. This leads to large end-to-end packet delays as depicted in~\ref{fig:cdf_del_100}. On the other hand, buffer aided relaying exploits the flexibility brought by the buffer in relay and utilizes channel variations efficiently. It transfers the data from BS buffer to relay buffer and from relay buffer to user, when the corresponding channels have good conditions, and therefore leads to low end-to-end delays. In particular, in this scenario, the average end-to-end packet delays are 22 ms and 1250 ms, respectively in buffer aided and conventional relaying.
 
The above results confirm that using buffer at relay, improves the throughput as well as the end-to-end delay in the system.

\vspace*{-4mm}
\section{Conclusion}\label{sec:conc}
In this letter we have studied the effect of buffering at relay on the end-to-end delay performance. Through the discussions about the queueing delay, we have explained the cause of delay in a simple queueing system. Based on that we have investigated the overall queueing delay in conventional and buffer aided relaying network and concluded that exploiting buffer at relay improves the system end-to-end delay. Using numerical results we have verified our discussions and shown that using buffers at relay leads to higher system throughput and lower end-to-end delay.
\bibliographystyle{IEEEtran}
\bibliography{IEEEabrv,refs}

\begin{thebibliography}{1}
\providecommand{\url}[1]{#1}
\csname url@samestyle\endcsname
\providecommand{\newblock}{\relax}
\providecommand{\bibinfo}[2]{#2}
\providecommand{\BIBentrySTDinterwordspacing}{\spaceskip=0pt\relax}
\providecommand{\BIBentryALTinterwordstretchfactor}{4}
\providecommand{\BIBentryALTinterwordspacing}{\spaceskip=\fontdimen2\font plus
\BIBentryALTinterwordstretchfactor\fontdimen3\font minus
  \fontdimen4\font\relax}
\providecommand{\BIBforeignlanguage}[2]{{%
\expandafter\ifx\csname l@#1\endcsname\relax
\typeout{** WARNING: IEEEtran.bst: No hyphenation pattern has been}%
\typeout{** loaded for the language `#1'. Using the pattern for}%
\typeout{** the default language instead.}%
\else
\language=\csname l@#1\endcsname
\fi
#2}}
\providecommand{\BIBdecl}{\relax}
\BIBdecl

\bibitem{JR:buffering3N}
B.~Xia, Y.~Fan, J.~Thompson, and H.~Poor, ``Buffering in a three-node relay
  network,'' \emph{IEEE Trans. Wireless Commun.}, vol.~7, pp. 4492-- 4496, Nov.
  2008.

\bibitem{JR:perf_rateless}
N.~Mehta, V.~Sharma, and G.~Bansal, ``Performance analysis of a cooperative
  system with rateless codes and buffered relays,'' \emph{IEEE Trans. Wireless
  Commun.}, vol.~10, pp. 2816 -- 2840, Jan. 2011.

\bibitem{JR:buf_fixedmixed}
N.~Zlatanov and R.~Schober, ``Buffer-aided relaying with adaptive link
  selection-fixed and mixed rate transmission,'' \emph{IEEE Trans. Inform.
  Theory}, vol.~59, pp. 2816 -- 2840, May 2013.

\bibitem{JR:buf_chal}
N.~Zlatanov, A.~Ikhlef, T.~Islam, and R.~Schober, ``Buffer-aided cooperative
  communications: Opportunities and challenges,'' \emph{IEEE Communications
  Magazine}, vol.~52, pp. 146-- 153, Apr. 2014.

\bibitem{book:stochasticopt}
M.~J. Neely, \emph{Stochastic Network Optimization with Application to
  Communication and Queueing Systems}.\hskip 1em plus 0.5em minus 0.4em\relax
  Morgan \& Claypool, 2010.

\bibitem{simulation}
{3GPP TR 25.996 V7.0.0 (2007-06), Tech. Rep}, ``Spatial channel model for
  multiple input multiple output (mimo) simulations.''

\end{thebibliography}

\end{document}